\documentclass[twocolumn, usenatbib]{aastex62}

\usepackage{graphicx}	
\usepackage{amsmath}	
\usepackage{amssymb}	

\received{\today}
\revised{}
\accepted{}

\shorttitle{Radio Nanoshots}
\shortauthors{Philippov, Uzdensky, Spitkovsky, Cerutti}

\begin{document}

\newcommand{\red}{\textcolor{red}}
\newcommand{\blue}{\textcolor{blue}}
\newcommand{\green}{\textcolor{green}}

\title{Pulsar Radio Emission Mechanism: Radio Nanoshots as a Low Frequency Afterglow of Relativistic Magnetic Reconnection}

\correspondingauthor{Alexander Philippov}
\email{sphilippov@flatironinstitute.org}

\author{Alexander Philippov}
\affiliation{Center for Computational Astrophysics, Flatiron Institute, 162 Fifth Avenue, New York, NY 10010, USA}

\author{Dmitri A. Uzdensky}
\affiliation{Center for Integrated Plasma Studies, Department of Physics, 390 UCB, University of Colorado, Boulder, CO 80309, USA}

\author{Anatoly Spitkovsky}
\affiliation{Department of Astrophysical Sciences, Peyton Hall, Princeton University, Princeton, NJ 08544, USA}

\author{Beno\^it Cerutti}
\affiliation{Univ.\ Grenoble Alpes, CNRS, IPAG, 38000 Grenoble, France}

\begin{abstract}
In this Letter we propose that coherent radio emission of Crab, other young energetic pulsars, and millisecond pulsars is produced in the magnetospheric current sheet beyond the light cylinder. We carry out global and local two-dimensional kinetic plasma simulations of reconnection to illustrate the coherent emission mechanism. Reconnection in the current sheet beyond the light cylinder proceeds in the very efficient plasmoid-dominated regime, and current layer gets fragmented into a dynamic chain of plasmoids which undergo successive coalescence. Mergers of sufficiently large plasmoids produce secondary perpendicular current sheets, which are also plasmoid-unstable. Collisions of plasmoids with each other and with the upstream magnetic field eject fast-magnetosonic waves, which propagate upstream across the background field and successfully escape from the plasma as electromagnetic waves that fall in the radio band. This model successfully explains many important features of the observed radio emission from Crab and other pulsars with high magnetic field at the light cylinder: phase coincidence with the high-energy emission, nano-second duration (nanoshots), and extreme instantaneous brightness of individual pulses.  
\end{abstract}

\keywords{keywords --- plasmas -- pulsars: general -- magnetic reconnection -- radiation mechanisms: non-thermal}

\section{Introduction and Radio Emission paradigm}
What causes pulsar radio emission has remained a mystery for nearly half a century. This is mainly because the radio emission is coherent, e.g.,  produced by microscopic magnetospheric collective plasma motions that are extremely difficult to calculate analytically. Recent discovery of coherent radiation from Fast Radio Bursts (FRBs) \citep{Lorimer07}  underscores the role of  pulsars  as  keys to understanding astrophysical coherent emission and warrants a reevaluation of the radio emission mechanisms with modern techniques. 

Pulsar magnetospheres have recently been  successfully studied numerically using first-principles particle-in-cell (PIC) approach. PIC simulations helped uncover the conditions for the operation and intermittency of the polar cap discharge \citep{Timokhin13,Chen14,Philippov15b}, located  high-energy emission source at the current sheet beyond the light cylinder (LC) \citep{Cerutti16}, and explained the general morphology of gamma-ray profiles \citep{Philippov18}. They also provided clues about potential mechanisms of coherent radio emission. First, non-stationary screening of the electric field by pair cascade at the polar cap could lead to the production of coherent electromagnetic (EM) waves (\citealt{Timokhin13,Philippov15b}, Philippov, Timokhin \& Spitkovsky, in prep). Second, the return separatrix current layer bounding the closed-field zone supports multi-streaming particle distributions \citep{Philippov18,Brambilla18}, which could be unstable to anomalous cyclotron instability (\citealt{Lyutikov99}, Philippov \& Lyutikov, in prep). Finally, magnetic reconnection in the current sheet beyond the LC proceeds in the highly non-stationary, plasmoid-dominated regime \citep{Cerutti17}. \cite{Uzdensky14} suggested that rapid coherent motions of plasmoids in the hierarchical chain could contribute to radio emission, albeit without proposing a concrete physical mechanism.

The global magnetospheric structure suggests that beamed emission at the tip of the separatrix layer maps into the same caustic as the current-sheet emission \citep{Bai10}, so the three possible emission mechanisms described above lead to two potential sources of radio emission: emission produced in the non-stationary polar discharge, which we call Type~I, and emission produced in the outer magnetosphere, Type~II. Type II emission should naturally coincide in phase with the magnetospheric gamma-ray emission, which also originates in the current layer \citep{Lyubarsky96, Uzdensky14, Cerutti16,Kalapotharakos18}. This emerging picture is motivated by the recent observations of the Crab pulsar, several other young energetic pulsars, and a few millisecond pulsars, which clearly show a radio emission source that is coincident with the location of the pulsed high-energy radiation source \citep{Johnston04,Johnson14,Eilek16}. 
In the Crab, this radio emission component (identified here with Type-II emission) is sometimes resolved into discrete bright pulses of nanosecond duration, \textquotedblleft nanoshots\textquotedblright \citep{Eilek16}. These nanoshots may entirely form the Crab's main pulse and low-frequency interpulse, and appear to come in packets making up microsecond-long bursts. 

In this Letter we present a physical model of coherent Type II radio emission produced by merging plasmoids in the reconnecting current sheet beyond the pulsar's~LC. In Section 2 we discuss the plasmoid instability in global PIC magnetospheric simulations. In Section 3 we present local PIC simulations of reconnection, which allow us to calculate the properties of coherent low-frequency EM radiation emitted by merging plasmoids. We summarize our model's observational implications and conclude in Section~4.

While this paper was in preparation, this idea has been independently suggested by Y.~Lyubarsky \citep{Lyubarsky19}. Our work complements his paper with simulations and shows that relativistic reconnection is indeed a powerful source of low-frequency waves, which for some pulsars are in the GHz band.

\section{Global magnetospheric simulations}
\begin{figure*}
  \begin{center}
    \includegraphics[scale=0.235]{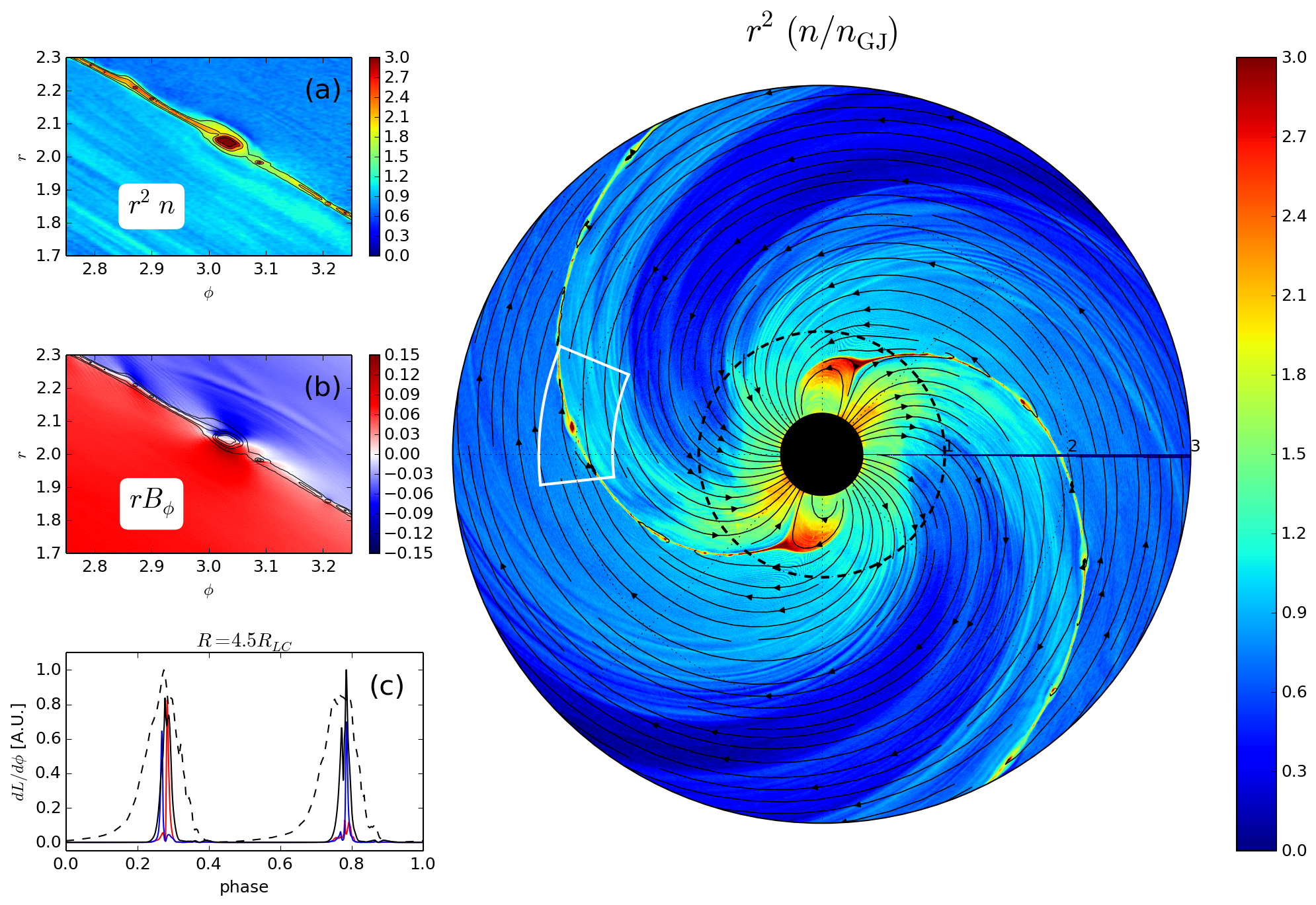}
  \end{center}
  \vspace{-5mm}
  \caption{ \label{fig:density_global} Plasma density in the 2D global simulation of the striped wind in the equatorial plane, for plasma multiplicity at the stellar surface $\kappa=100$. Current sheet beyond the LC is unstable to tearing instability and is fragmented into plasmoids. The black dashed circle shows the pulsar's LC. Sub-panels show the zoom-in into the plasmoid chain in a region highlighted with a white rectangle and present (a) the plasma density, and (b) the toroidal component of the magnetic field. Panel (c) shows the lightcurve of two radio pulses (blue and red lines), the radio lightcurve averaged over two rotational periods of the pulsar (black solid line) and the gamma-ray profile (black dashed line).}
  \vspace{2mm}
\end{figure*}

Numerical simulations of pulsar magnetospheres, starting with the pioneering work of \citet{Contopoulos99}, showed the formation of a prominent current sheet beyond the~LC. Global two-dimensional (2D) \citep{Cerutti17} and three-dimensional (3D) PIC simulations \citep{Philippov15,Cerutti16,Philippov18} also confirmed that this current sheet is unstable to plasmoid instability. However, because of their limited scale separation between the current sheet's length, comparable to the LC radius ($R_{\rm LC}$), and its width, $\delta\sim r_L$ (here, $r_L$ is the average Larmor radius of particles in the sheet), 3D simulations typically show only 1-2 plasmoids. In contrast, 2D simulations of the striped wind that are limited to the equatorial plane allow significantly greater scale separation and hence show more plasmoids \citep{Cerutti17}.

To probe the expected number of plasmoids in the current sheet in the global pulsar magnetosphere as a function of plasma parameters, we simulate the structure of the striped wind using the setup of \cite{Cerutti17}. Our simulations are performed in spherical coordinates $(r,\theta,\phi)$, and are restricted to the equatorial plane ($\theta=\pi/2$). Pulsar's LC is at $r=R_{\rm LC}=3R_*$, where $R_*$ is the stellar radius. We inject neutral plasma at the pulsar  surface and give it a nonrelativistic kick along the magnetic field. We run this setup for several values of the plasma multiplicity, $\kappa=1,10,100$, at 
$r=R_*$. Multiplicity is defined as $\kappa=n/n_{\rm GJ}$, where $n$ is the injected plasma density and $n_{\rm GJ}={\Omega}_*B/2 \pi c e$ is the fiducial Goldreich-Julian plasma density (here, $\Omega_*$ is the stellar angular velocity and $B$ is the local magnetic field strength). We choose the magnetic field at the star to be sufficiently strong so that even the simulation with the largest injected multiplicity, $\kappa=100$, has a sufficiently high value of the magnetization parameter at the LC, $\sigma_{\rm LC} \equiv B^2_{\rm LC}/4\pi \kappa n_{\rm GJ} m_e c^2\approx 70$ (hereafter, $B_{\rm LC}$ is the magnetic field strength at the~LC). Our simulations include the synchro-curvature radiation reaction on particle motion (see \citealt{Cerutti17} for details).

Simulations quickly settle to a solution with closed and open field lines, and a current sheet beyond the LC, separating regions of oppositely directed magnetic field. Typical structure of the plasma density is shown in Figure~\ref{fig:density_global}. We find that the current sheet width, $\delta$, scales with the Larmor radius of particles heated up to energies $\langle\gamma\rangle\simeq \sigma_{\rm LC}$, so that $\delta \propto r_L \propto m_e c^2 \sigma_{\rm LC} / e B_0 \propto R_{\rm LC} {\kappa}^{-1}$.  For our $\kappa=1$ simulation, the ratio $\delta/R_{\rm LC}$ is substantial, $\delta/R_{\rm LC} \approx 0.15$, and the sheet is stable to tearing. Simulation with $\kappa=10$ develops a current sheet with $\delta/R_{\rm LC} \approx 0.026$ and shows few plasmoids, while the $\kappa=100$ simulation has $\delta/R_{\rm LC} \approx 0.0028$ and develops a vigorous plasmoid chain (see Figure~1). The zoom-in into the chain at $2 R_{\rm LC}$ is shown in Figure~\ref{fig:density_global}a. Overall, we find that the number of plasmoids within $2R_{\rm LC}$ scales approximately as $R_{\rm LC}/\delta$, but we caution that our dynamic range is too small to predict an asymptotic scaling. However, it is clear that for real parameters of the Crab, this would correspond to very large numbers of plasmoids between one and two~$R_{\rm LC}$. We also observe plasmoid mergers at $r \lesssim (1-3) R_{\rm LC}$, with most frequent mergers of small plasmoids occurring close to the~LC.  As plasmoids grow when they propagate away from the LC, we observe mergers of bigger plasmoids at larger distances. We note that mergers of all sizes should be much more frequent for realistic parameters. Beyond $r\approx 4 R_{\rm LC}$ the flow expands nearly radially, and the plasmoid structure freezes out. 

Figure~\ref{fig:density_global}b shows the toroidal magnetic field component, exhibiting a wave-like structure (most clearly seen around $r \approx 2.2 R_{\rm LC}$, $\phi \approx 3.1$) coming from big plasmoids. As discussed below, these waves are fast waves emitted in the process of plasmoid mergers. We compute the lightcurve of this low-frequency radiation by measuring the Poynting flux at the simulation boundary, $r=6R_{\rm LC}$ and $\phi=0$, and subtracting the \textquotedblleft DC\textquotedblright components of the EM field which correspond to the bulk Poynting flux in the pulsar wind. The result is a sequence of bright short pulses with variable arrival times clustered around rotational phases 0.3 and 0.8 (see Figure~\ref{fig:density_global}c). The lightcurve averaged over a few pulsar rotations is shown as a black solid line; it is significantly narrower than the lightcurve of the incoherent high-energy synchrotron emission of particles accelerated in the current sheet (black dashed line, reproduced from \citealt{Cerutti17}). As the synchrotron emission is in the gamma-ray band for young energetic pulsars \citep{Uzdensky14,Hakobyan18}, we find that the peaks of coherent low-frequency emission coincide with the peaks of the gamma-ray lightcurve\footnote{In the Crab, radio and gamma-ray peaks are slightly offset in phase \citep{Abdo10}. 3D magnetospheric modeling for realistic inclination and viewing angles is required to reproduce observed lightcurves.}. Below we argue that the characteristic frequency of the low-frequency emission from plasmoid mergers falls in the GHz range for parameters of the Crab and other pulsars with high $B_{\rm LC}$.

\section{Local simulations}

\begin{figure}
  \begin{center}
    \includegraphics[scale=0.55]{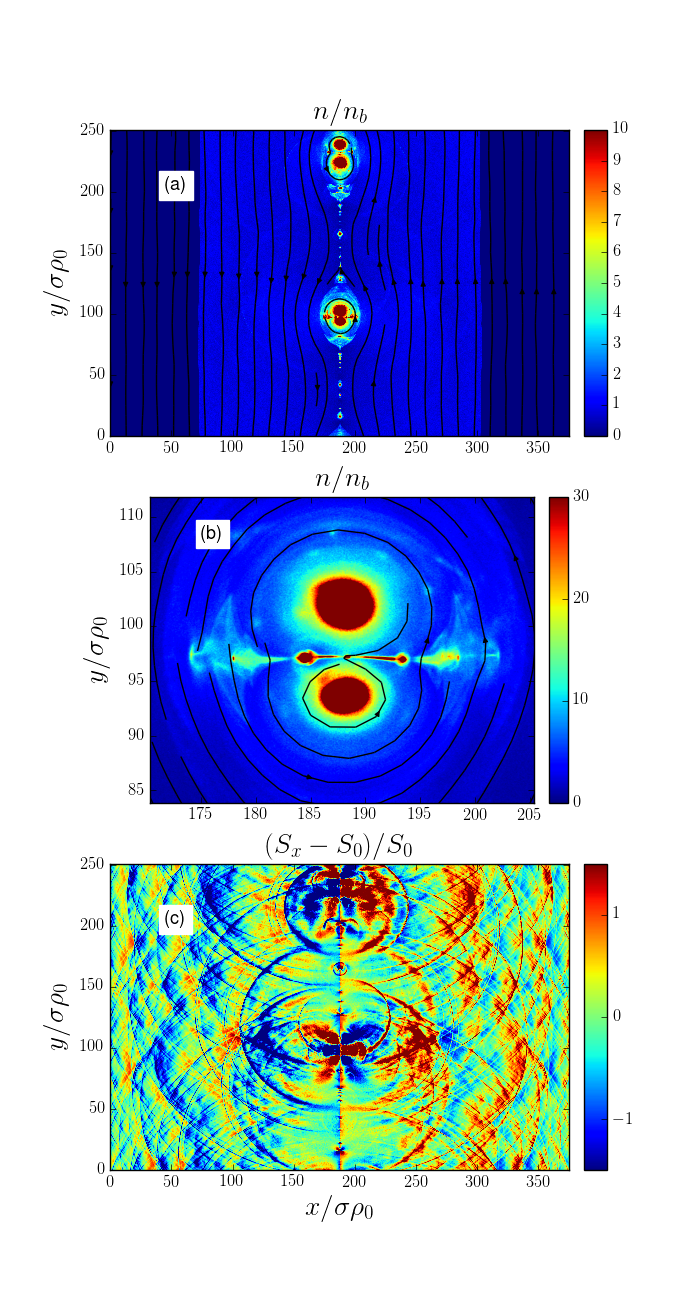}
  \end{center}
  \vspace{-5mm}
  \caption{ \label{fig:poynting} (a): Snapshot of the plasma density in the local simulation showing the current sheet breaking into a chain of plasmoids. (b): A zoom-in into a merger of two big plasmoids, showing  that the perpendicular current sheet at the plasmoid interface also breaks and produces secondary plasmoids. (c): The $x$-component of the Poynting flux in the simulation, with the reconnection-induced Poynting flux $S_0=0.1 c B^2_0/4\pi$ subtracted. Spikes of Poynting flux produced by plasmoid mergers escape from the reconnection layer.}
  \vspace{2mm}
\end{figure}

\subsection{Numerical setup}
To understand the emission mechanism, we run local 2D PIC simulations of relativistic reconnection using Tristan-MP code \citep{Spitkovsky05}. 
We adopt a standard setup with one relativistic Harris pair-plasma current sheet \citep{Kirk03}, plus a uniform cold background pair plasma 
of total density~$n_b$. The box dimensions are $L_x \times L_y$  ($L_x = 1.5L_y$), with $y$ parallel to the reconnecting magnetic field $B_0$ and $x$ perpendicular to the current sheet; $z$ (not simulated) parallels the initial current. Our boundary conditions are periodic in the $y$ direction. In the $x$ direction we introduce a buffer zone at $x=[-L_x,-0.8L_x]$ and $x=[0.8L_x, L_x]$ with no plasma particles, and with a radiative boundary condition for both fields and particles at the edge of the simulation box. These boundary conditions are new in reconnection studies, and are specifically designed to study how the waves produced by the coherent motions of plasmoids escape from the plasma into surrounding vacuum. We do not add any guide field because it should be negligible in pulsar current sheets. We vary the upstream magnetization parameter, $\sigma \equiv B^2_0/(4\pi n_b m_e c^2)$, from 20 to 100 in our series of simulations, but we find that the resulting  radio emission shows almost no $\sigma$-dependence. Our system size, $L_y = 250\rho_0$ (where $\rho_0 \equiv m_e c^2/eB_0$ is a nominal cold relativistic gyroradius), allows us to observe multi-scale plasmoid chain formation and many subsequent dynamical mergers.

\subsection{Results}
Reconnection starts as the tearing instability breaks up the current layer into a chain of plasmoids. As these primary plasmoids move along the sheet, secondary current sheets between them become sufficiently elongated and also become unstable, yielding a hierarchical fragmented structure (see Figure~2a for the snapshot of magnetic field lines and plasma density). Over time, plasmoids grow and merge into larger ones. Current sheets perpendicular to the main current sheet form between sufficiently large merging plasmoids. These sheets also become unstable to secondary tearing and form secondary plasmoids as shown in the density map in Figure~2b. Both small plasmoids in the main sheet and plasmoids in the secondary perpendicular current sheets can reach relativistic velocities. As they merge with each other or collide with the upstream magnetic field (as is the case for plasmoids in secondary perpendicular current sheets), coherent time-dependent currents are generated at the interface and launch a powerful EM wave that lasts until the merger is complete. These waves are fast magnetosonic waves, which in the high-$\sigma$ limit become vacuum EM waves. They propagate upstream across magnetic field and are unaffected when they cross the buffer transition at $x=0.2L_x$ or $x=0.8L_x$. The electric field of the waves is then recorded at the edge of the simulation domain, $x=0$ and $x=L_x$. 

\begin{figure}
  \begin{center}
    \includegraphics[width=8.6cm, trim = 3.0mm 2.0mm 2.5mm 2.0mm, clip]{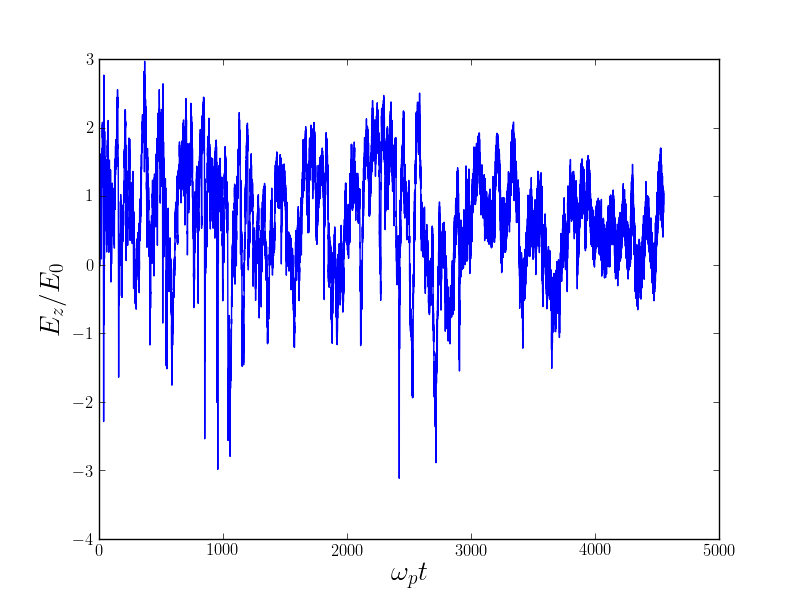}
    \includegraphics[width=8.6cm, trim = 3.0mm 2.0mm 2.0mm 10.0mm, clip]{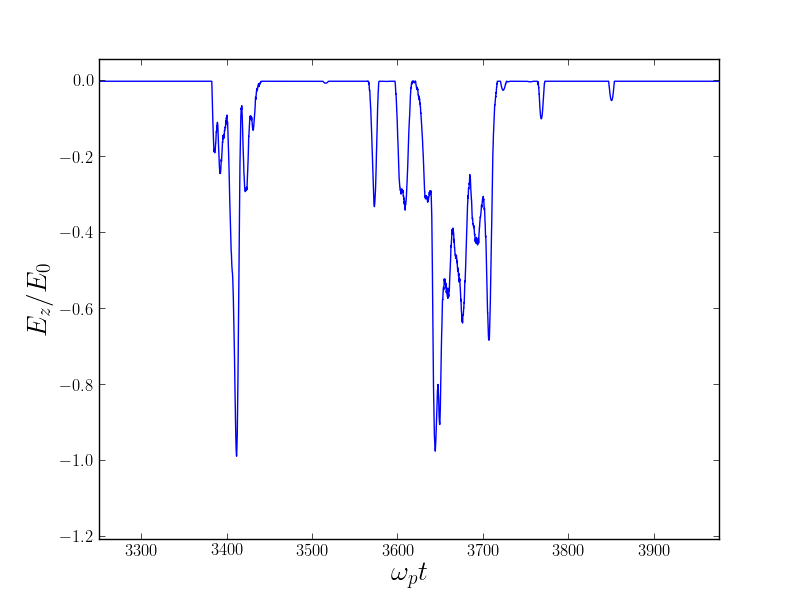}
  \end{center}
  \vspace{-5mm}
  \caption{\label{fig:ez} Top panel: Electric field calculated at the boundary of the simulation box, $x=0,y=L_y/2$. Bottom panel: A zoom-in into a shorter time interval: only negative values of the electric field, corresponding to outgoing waves, are kept for clarity.}
  \vspace{-2mm}
\end{figure}

In Figure \ref{fig:poynting}c we show a map of the $x$-component of the Poynting flux in a simulation snapshot, where we subtracted the average reconnection-driven inflowing Poynting flux, $S_0\approx 0.1 c B^2_0/4\pi$. The circular spikes correspond to outward-propagating EM waves generated in plasmoid mergers and in collisions of plasmoids with the upstream field. Each individual spike is a half-wavelength oscillation, so that the spike duration~$\tau$ and the associated frequency~$\nu$ are related as $\tau\nu \simeq 1$. 

\begin{figure}
  \begin{center}
    \includegraphics[width=8.6cm, trim = 3.0mm 2.0mm 2.5mm 2.0mm,clip]{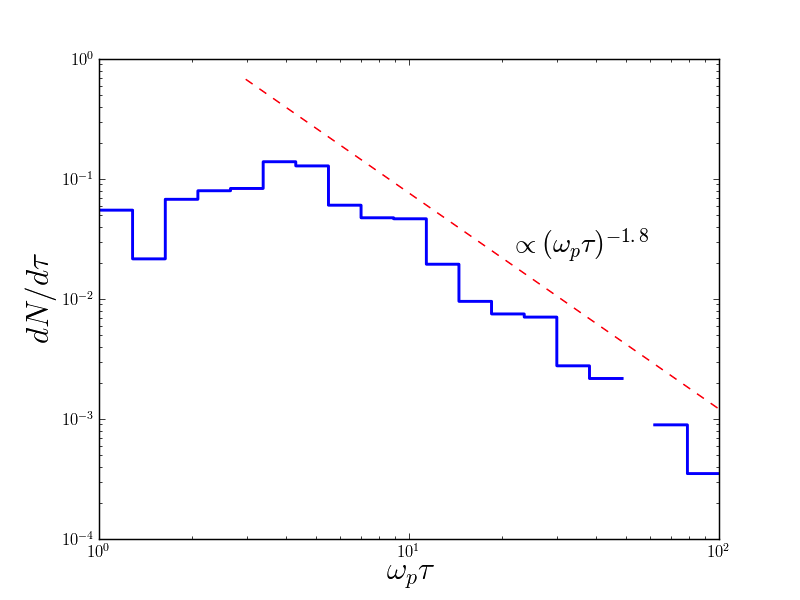}
  \end{center}
  \vspace{-5mm}
  \caption{ \label{fig:hist} Distribution of single spikes and pulses over their durations.}
  \vspace{2mm}
\end{figure}

Figure \ref{fig:ez} shows the evolution of the electric field, normalized by the reconnection electric field $E_0=\langle E_z \rangle_y \simeq 0.1 B_0$, at the $x=0$ boundary of the box at $y=L_y/2$. Here, regions of $E_z<0$ correspond to waves that carry energy away from the layer. In the bottom panel we show a zoom into a short time interval, when the waves from several different mergers reach the simulation boundary (only negative values of $E_z$ shown for clarity). The pulse, e.g., a train of several spikes, between $t=3600-3800\omega_p t$ comes from a merger of two big plasmoids, which produced several smaller plasmoids in the perpendicular sheet between them. Each individual merger of a small plasmoid with the upstream field generates a single electric-field spike. The overall pulse duration is $\tau\approx 10/\nu$, where $\nu$ is the frequency associated with individual spikes. This relation is consistent with the observed duration of individual radio nanoshots from the Crab \citep{Eilek16}.

The duration distribution of single spikes and pulses\footnote{We do not distinguish between pulses and spikes to calculate this distribution. We note that individual spikes occur significantly more often in our simulation setup.} is shown in Figure~\ref{fig:hist}. It peaks around $\tau\sim 10/\omega_p$, where $\omega_p$ is the background plasma frequency, and resembles a power law ${\rm d} N/{\rm d}\tau \approx (\omega_p \tau)^{-\alpha}$ for larger durations, and $\alpha$ is between 1.5 and 2. The pulse-averaged spectral density of the EM radiation in 3D can be then estimated as $S_{\nu}{\rm d}\nu \propto \left(\mathcal{E}/\tau\right) \left({\rm d}N/{\rm d}\tau\right) {\rm d} \tau \propto \tau^{2-\alpha} {\rm d} \tau$, where we used  $\tau \nu \simeq 1$, and estimated the energy of the individual pulse, $\mathcal{E}$, as the magnetic energy stored in the corresponding merging plasmoid of size $\ell$, which scales as the plasmoid's volume $\sim {\ell}^3\sim (c\tau)^3$. The obtained steep spectrum, $S_{\nu}\propto\nu^{-\beta}$, where $\beta=(4-\alpha)$ is between 2 and 2.5, implies that most of the energy is emitted as low-frequency radiation, produced by mergers of moderate-size plasmoids. This spectrum is compatible with the observed $\beta\approx 3$, for the main component of the Crab's radio profile \citep{Moffett99}, although the exact value of $\beta$ varies substantially between the pulses \citep{Karuppusamy10}.

While the occasional brightest spikes have instantaneous amplitudes of up to~$3 E_0$, the electric field in typical pulses is about~$E_0$, implying typical instantaneous Poynting fluxes of the EM waves as high as $c E_0^2/4\pi \approx 0.01 c B_0^2/4\pi$. However, when averaged over the simulation duration~$T$, the Poynting flux of the outgoing EM wave emission, measured at $x=0$ boundary of the box at $y=L_y/2$, is much smaller, $(c/4\pi)\int_{0}^{T}E_z(B_y-\langle B_y\rangle_y) {\rm d} t /T \approx 10^{-4} c B_0^2/4\pi$. This implies that only a small fraction of the reconnection-dissipated power is transferred into coherent EM waves. Most of this coherent power is radiated at low frequencies, which may not necessarily fall in the observed GHz range in the case of pulsars. Thus, the conversion efficiency into observable radio waves could be even lower. We note that the average efficiency in our simulations is likely significantly exaggerated, since the periodic boundary conditions produce favorable conditions for plasmoid mergers. Mergers are still quite regular in simulations with outflowing boundaries \citep{Sironi16}; however, it appears hard to quantify the average efficiency of the process in real pulsars with our idealized local calculations. Although the basic picture of the current sheet breaking into plasmoids as a result of tearing instability stays the same in three dimensions \citep{Werner17}, our quantitative conclusions remain to be checked with future 3D simulations. Nonlinear interactions between the fast waves may further decrease the average efficiency of the escaping radiation \citep{Lyubarsky19}.

\section{Observational Implications and Conclusions}

We now discuss implications of our results for observations of pulsar radio emission. In the Crab magnetosphere, the comoving plasma density and temperature inside the reconnecting current sheet are regulated by cross-layer pressure balance and the energy balance between reconnection-powered  heating and synchrotron cooling, to be about $10^{13}\, {\rm cm}^{-3}$ and 10~GeV, respectively (\citealt{Lyubarsky96,Uzdensky14}). Then, the typical thickness $\delta$ of elementary current layers is of order 1 meter \citep{Uzdensky14,Lyubarsky19}, and the typical size of small plasmoids, about 10-100 times larger, is of order 30 meters. This translates in a comoving duration of coherent radio-spikes of about 100~ns. After accounting for relativistic Doppler effect due to pulsar-wind bulk Lorentz factor $\Gamma \sim 10-100$ (see, e.g., \citealt{Lyubarsky96}), this can explain the observed GHz-range of pulsar radio emission \citep{Uzdensky14,Lyubarsky19} and, in particular, ns-scale pulses (nanoshots). Pulsars with $B_{\rm LC}$ weaker than in the Crab have thicker current sheets and hence larger secondary plasmoids. Thus, coherent EM emission from plasmoid mergers in these systems may occur at frequencies lower than the traditional $\gtrsim 100$~MHz radio range, of interest for possible detections with LOFAR and SKA. This may explain why radio emission phase-aligned with high-energy emission is only observed in pulsars with high~$B_{\rm LC}$ \citep{Johnston04}. 

It is remarkable that a nanosecond-long radio spike produced from a region just tens of meters across in a single plasmoid merger in the magnetosphere of Crab pulsar 2~kpc away is so powerful that it can be detected at Earth. This is perhaps the smallest-size source of observable emission in all of astrophysics. Indeed, the amount of magnetic energy released in an individual merger event is about $\mathcal{E} \sim (B_{\rm LC}^2/8\pi) \ell^3 \sim 10^{21}\, {\rm erg}$, where we took  $B_{\rm LC} = 1~{\rm MG}$ and $\ell=30$~m. If, say, 1\% of this energy is emitted as coherent EM pulse with a duration $\tau$ of a few ns-long wave-periods (i.e., $\tau \nu \sim {\rm few}$), one can estimate the observed spectral flux density (for a distance of $d=2$~kpc) of order 
$S_{\rm obs} \sim 0.01 (\pi d^2)^{-1} (\mathcal{E}/\tau\nu) \Gamma^3 \sim  10^{-25}\Gamma^3 {\rm erg/(cm^2 \cdot sec \cdot Hz)} \sim 0.01\,\Gamma^3 \,{\rm Jy}$; taking $\Gamma \sim 30$, we get $S_{\rm obs} \sim 300 \,{\rm Jy}$, comparable to the observed values (\citealt{Eilek16}, see also \citealt{Lyubarsky19}). Here we assumed that this emission is isotropic in the pulsar wind frame, which is the case in our idealized two-dimensional reconnection simulation. The corresponding brightness temperature of 0.3~GHz radio emission is about $T_B \sim S_{\rm obs}c^2/(2k_B \nu^2\delta\Omega)\sim 10^{38}\, {\rm K}\, (\Gamma/30)^3$, where $\delta\Omega\sim \pi\ell^2/d^2$ is the solid angle covered by the colliding plasmoid, which is well in line with the observed values \citep{Hankins03}.

To conclude, our model of the radio nanoshots explains their observed frequency, duration, phase-alignment with the high-energy emission, extremely high instantaneous brightness and low average efficiency. Our simulations do not address the sometimes strong circular polarization of nanoshots \citep{Eilek16}. The observed circular polarization may arise due to propagation effects in the magnetospheric plasma \citep{Lyubarsky99,Beskin12}, which for high-$B_{\rm LC}$ pulsars imply the polarization formation radius located beyond the~LC.

Given the right physical conditions, our results may have implications for coherent radio emission from other sources where relativistic reconnection may take place, e.g., magnetar magnetospheres and interacting magnetospheres of binary neutron stars before the merger. These possibilities and potential connections to FRBs will be investigated elsewhere. 

We thank V.~Beskin and Y.~Lyubarsky for stimulating discussions. The Flatiron Institute is supported by the Simons Foundation. This research was supported by the NSF (grant AST-1411879), NASA ATP (grants NNX16AB28G, NNX17AK57G, and 80NSSC18K1099), CNES and the Universit\'e Grenoble Alpes (IDEX-IRS grant). DU acknowledges the hospitality of the Institute for Advanced Study and the Ambrose Monell Foundation for supporting his 2016-2017 sabbatical, when most of this work was done. We acknowledge PRACE for awarding us access to Curie at GENCI@CEA, France.

\end{document}